# A High-resolution Large-eddy Simulation Framework for Wildland Fire Predictions using TensorFlow

Qing Wang[1*], Matthias Ihme[1,2], Rod R. Linn[3], Yi-Fan Chen[1], Vivian Yang[1], Fei Sha[1], Craig Clements[4], Jenna S. McDanold[3], John Anderson[1]

[1]Google, 1600 Amphitheatre Parkway Mountain View, CA 94043, USA

[2]Stanford University, 440 Escondido Mall, Stanford CA 94305, USA

[3]Los Alamos National Laboratory, Bikini Atoll Rd, Los Alamos, NM 87545, USA

[4]San Jose State University, 1 Washington Square, San Jose, CA 95192, USA

**Background**

As the impact of wildfires has become increasingly more severe over the last decades, there is continued need for improvements in our ability to predict wildland fire behavior over a wide range of conditions and spatial scales. One approach towards this goal is through coupled fire/atmosphere modeling tools. While significant progress has been made on advancing their physical fidelity, existing modeling tools have not taken full advantage of emerging programming paradigms and computing architectures to enable high-resolution wildfire simulations.

**Aims**

[1*] Corresponding author.
  E-mail address: wqing@google.com (Q. Wang).

This work presents a new wildfire simulation framework that enables landscape-scale wildfire simulations with physical representation of the combustion at an affordable computational cost.

**Methods**

We developed a coupled fire/atmosphere simulation framework using TensorFlow, which enables highly efficient and scalable computations on Tensor Processing Unit (TPU) hardware architecture.

**Key Results**

Simulations of the prescribed fire experiment FireFlux II (Clements et al., 2019) are performed with different mesh resolutions (up to 0.5 m horizontal and 0.125 m vertical resolutions) for validation and fire behavior analysis.

**Conclusions**

A parametric study on the mesh resolution shows that the global quantities, such as the fire-scar area and fire-spread rate, are fairly insensitive to the horizontal mesh resolution within a range between 0.5 m and 2 m, which is sufficient for predicting dynamic fire properties associated with fine-scale turbulent structures in the atmospheric boundary layer.

**Implications**

This new simulation framework is efficient in capturing both global quantities and unsteady dynamics of wildfires at high spatial resolutions.

*Keywords: Wildfire modeling, TensorFlow, Fire/atmospheric coupling, Large-eddy simulation, Tensor processing units, Fire propagation*

## Introduction

The frequency and severity of wildfires has increased profoundly over the past decades (Westerling et al., 2006; Jolly et al., 2015; Abatzoglou & Williams, 2016) because of several confounding factors that include changes in fire suppression and fire management, extent of the wildland-urban interface, and climate. Many regions are experiencing extended fire seasons and increased annual burn area (Parks & Abatzoglou, 2020;

Westerling, 2016). The increased fire severity causes significant ecological and economical losses (Thomas et al., 2017) and health burdens (Burke et al., 2021), which creates challenges for environmental planning and fire management (Hessburg et al., 2021).

To address the need for predictive methods that can guide fire management, enable landscape management, inform policy decisions, and support scientific inquiry, simulation techniques of varying physical fidelity, computational complexity, and accuracy have been developed (Sullivan, 2009a, 2009b, 2009c). Accurately predicting wildfire dynamics across a wide range of wildland fire conditions requires the consideration of complex fire-atmosphere interactions, including the coupling between meteorology, heat transfer, turbulence, and combustion (Liu et al. 2019). To capture these coupling effects, physics-based models are needed (Bakhshaii & Johnson, 2019).

Physics-based simulation tools are distinguished into two major categories based on scale (Bakhshaii & Johnson, 2019). One of the categories focuses on predicting meso-scale wildland fire behavior. By integrating empirical (Rothermel, 1972) or algebraic physics-inspired (Balbi et al., 2009) fire-spread models into numerical weather prediction models, these simulation approaches enable the prediction of fire-spread behavior on scales spanning more than 100 km in real time. Examples of this category of models include WRF-SFIRE (Mandel et al., 2011), CAWFE (Coen, 2013), WRF-Fire (Coen et al., 2013), and MesoNH-ForeFire (Filippi et al., 2013). While these approaches have been shown to capture the global fire behavior, such as the rate of spread (ROS), their coarse spatial resolution limits the accurate prediction of fire intermittency and turbulent fire dynamics, which is critical for simulating unsteady fire events (Finney et al., 2015; Viegas et al., 2022).

The second category of physics-based simulation approaches focuses on fires at the micro-scale. In these approaches, the hydrodynamics is represented with large-eddy simulations (LES) or Reynolds-averaged Navier-Stokes (RANS) simulations, and is fully coupled with the fire dynamics through the use of a fire model that represents the chemical reactions. Representative models of this category are the CNRS Fuel Beds Simulator (Porterie et al., 2000), FIRETEC (Linn et al., 2002), WFDS (McGrattan et al., 2006), FireFOAM (Wang et al., 2011), and FireStar3D (Morvan et al., 2018). Although these models capture the unsteady fire-atmosphere interactions, their high computational cost has constrained the spatial extent and duration of fire predictions (Sullivan, 2009; Coen et al., 2020). As such, the use of these models in predicting the behavior of large wildfires remains challenging.

With the recent advancements in computer technologies, including hardware architectures and software stacks, opportunities arise to significantly improve the efficiency of physics-based wildfire simulations. Therefore, the objective of this work is to develop and validate a physics-based coupled fire/atmosphere solver, named SWIRL-FIRE, for large-scale high-fidelity wildland fire simulations. This solver is based on the open-source low-Mach hydrodynamic simulation framework SWIRL-LM (Wang et al., 2022a), and integrates sub-models of FIRETEC (Linn & Cunningham, 2005) to represent the combustion and multi-phase interaction with vegetation, thereby accounting for the coupling between the fire and the atmosphere. This simulation framework leverages the TensorFlow programming paradigm (Abadi et al., 2016) and tensor processing units (TPUs), which were developed for machine learning (ML) and scientific computing (Jouppi et al., 2021). While TensorFlow enables ML capabilities, the focus of the present study is on evaluating the physical model implementation and spatial resolutions; the utilization of ML will be addressed in future work. Access to this TPU computing hardware is available through the Google Cloud Platform, and the code is publicly available through GitHub (Wang et al., 2022a). The accelerated numerical computations by the new hardware allow us to simulate meso-scale wildfires at homogeneous spatial resolution of $O(10^{-1})$ m with $O(10^9)$ m grid points at an affordable operation cost. We evaluate this modeling tool in application to a prescribed fire experiment, FireFlux II (Clements et al., 2013), with specific attention paid to the effect of spatial resolution on representing the fire dynamics. The importance of mesh resolution and domain size has been discussed in previous studies (Moinuddin et al., 2018), showing that local refinement of the burnable fire region is necessary to obtain converged fire-spread predictions.

The remainder of this manuscript has the following structure. The next section presents methods that are used to construct the simulation framework, including the mathematical model, solution algorithm, and implementation. The experiment and computational setup section introduces the simulation setup based on the FireFlux II experiment. Results are presented in the subsequent section, discussing comparisons with experimental data and sensitivity analysis with respect to the mesh resolution. Concluding remarks are provided in the last section.

## Methods

*Mathematical model and solution algorithm*

In this work, we model the spatio-temporal dynamics of wildland fires using an LES approach, in which the governing equations for the gas-phase are described by the solution of the Favre-filtered conservation equations for mass, momentum, oxygen mass fraction, and potential temperature. To account for subgrid-scale effects, arising from the turbulent stresses and turbulent scalar transport, we employ the Smagorinsky-Lilly model (Lilly, 1962). The combustion of the solid fuel is modeled by a one-step mixing-limited oxidation reaction, and energetic impacts of moisture evaporation are accounted for. The implementation of the combustion model follows the work by Linn et al. (2002) and is summarized in Appendix A. Different from FIRETEC, which adopts a fully compressible formulation, the present work solves the governing equations with a low-Mach number approach with prescribed hydrostatic reference state. This enables advancing the solution at larger time-step sizes by removing the dependence on the acoustic wave propagation.

The governing equations are solved on a three-dimensional Cartesian coordinate system using a finite-difference discretization. All spatial operators are discretized on a collocated mesh, and time-staggering is employed for the temporal discretization. Along each direction the mesh is equidistant. The coupled system of equations is solved using a predictor-corrector method with a time-explicit iterative scheme. Further details on the numerical discretization, solution algorithm, and convergence properties can be found in Wang et al. (2022a).

*TPU computing architecture and implementation*

All simulations presented in this work are performed on the TPUv4 computing architecture. Each chip in this computing architecture has two tensor compute cores that are optimized for dense linear algebra operations, such as matrix-matrix multiplications that are performed by a specially designed processing unit called matrix-multiplication unit (MXU), providing a peak throughput of 275 teraflops per chip. Each chip has 32 GiB high-bandwidth memory for fast memory access. A total of 4096 chips (8192 cores) are connected through a high-speed, three-dimensional toroidal network for low-latency data communication between cores. The TPUv4 architecture supports 8-bit arithmetic natively, and high-precision arithmetic is available through software emulations. Simulations in the present study are performed with single precision for the desired balance between computational efficiency and numerical accuracy. We implemented the simulation framework with TensorFlow, which offers specific advantages by utilizing existing libraries and the just-in-time compilation to optimize the performance on TPU architectures (Wang et al., 2022a).

## Experiment and computational setup

To evaluate the accuracy and performance of the proposed simulation framework, we consider the FireFlux II experiment as a benchmark configuration. In the following, we summarize the experiment and describe the computational setup that is employed to simulate this experiment.

### *Experimental setup*

The FireFlux II experiment (Clements et al., 2019) is a prescribed grassland fire on a flat-field prairie site. This experiment included both ambient fuel and wind observations as well as fire-spread. Based on atmospheric measurements prior to ignition, a west-northwest wind with an average 10-m wind speed of 8.5 m/s was reported. The mean temperature was 16 ℃. The unit-average fuel loading was reported as 0.64 kg/m$^2$ with a variation between 0.41 kg/m$^2$ and 0.81 kg/m$^2$ due to fine-particle heterogeneities from the upper- and lower-layer grasses, forbes, and shrubs.

The field was instrumented to simultaneously measure the development of the head fire and atmospheric conditions. This instrumentation included meteorological towers for locally measuring near-surface momentum, heat transport, wind velocity, and sonic temperature. Discrete measurements were performed at the main tower at measurement heights of 5.8, 10, 20, and 43 m above the ground level (AGL). The fire spread was measured using a total of 20 thermocouple-based HOBO temperature data loggers placed at the soil surface and arranged on a grid to record the fire-spread rate. Further details on the measurements, instrumentations, and data acquisition can be found in Clements et al. (2019). Available measurements are used to compare our simulation results.

*Computational setup*

In the present work, we consider experimentally reported conditions of the FireFlux II experiment, and no attempts were made to tune model parameters or operating conditions to match experimental results. The conditions and properties used in our simulations are summarized in Table 1.

Table 1. Physical parameters in the simulations.

| Operating condition [unit] | Value |
| --- | --- |
| 10-m wind speed (free stream) [m/s] | 8.5 |
| Wind direction (with respect to y-direction) [°] | -10 |
| Fuel type | Tall grass |
| Unit fuel load [$kg/m^2$] | 0.6 |
| Fuel distribution | Homogeneous |
| Fuel height [m] | 1.5 |
| Moisture content [%] | 14.2 |

To represent the fire spread over the course of the experiment, we adopted a prismatic domain with size $1000 \times 500 \times 1200$ m$^3$ in streamwise ($x$), spanwise ($y$), and vertical ($z$) directions, respectively. The streamwise direction is aligned with the mean wind from west-northwest direction. The height of the domain is selected to capture the atmospheric boundary layer. A parametric study was performed by increasing the lateral extend of the domain to 1000 m, showing no appreciable changes in the fire behavior.

To represent the shear stress and heat flux of the atmospheric boundary layer, the bottom-surface boundary is modeled with the Monin-Obukhov similarity theory (Stoll & Porté-Agel, 2006). A Rayleigh damping layer (Klemp & Lilly, 1978) is used within the top 10% of the computational domain. For a generic variable $\phi$ with a background value $\phi_0$, the Rayleigh damping term is computed as $f_{\phi,RD} = -\beta_{RD}(\phi - \phi_0)$, where $\beta_{RD}$ is a relaxation coefficient, which is a function of the depth ($\zeta$) into the total thickness of the Rayleigh damping layer ($\zeta_0$). In our simulations, $\beta_{RD}(\zeta) = (20\Delta t)^{-1} \sin^2\left(\frac{\pi}{2}\frac{\zeta}{\zeta_0}\right)$ (Klemp & Lilly, 1978), with $\zeta_0 = 0.1 l_z$ and $\zeta \in [0, \zeta_0]$, and $\Delta t$ is the step size in the simulation. This model is employed to prevent entrainment into the flow domain or the generation of unphysical flow structures at the top of the computational domain. Along the lateral sides of the computational domain, adiabatic and free slip boundary conditions are enforced.

Boundary conditions at the inflow are prescribed by a time-dependent turbulent inflow profile. These inflow conditions were obtained by performing an auxiliary flow simulation in the computational domain without igniting the fire, assuming that the atmospheric boundary layer is neutrally stratified. To obtain a fully developed turbulent boundary layer, we simulate a temporally evolving boundary layer by employing periodic boundary conditions in streamwise direction (Linn et al., 2013). To match the experimental condition, the free-stream wind speed is set to the experimentally reported 10-m wind of 8.5 m/s. After advancing this simulation over 100 flow-through times (FTTs), defined as $t_{FTT} = l_x/U_\infty$ (with $l_x$ being the length of the domain and $U_\infty$ the free-stream wind speed), to establish a statistically stationary turbulent boundary-layer profile, we extract the three-dimensional velocity field at the *y-z* cross-section 500 m from the inlet of the domain for 500 s, corresponding to 4.25 FTTs. An analysis of the atmospheric boundary layer is presented in Fig. A1.

To perform the fire simulations, we initialized the flow field with the last snapshot from the auxiliary turbulent inflow simulation. The potential temperature profile that is used to initialize the flow field was obtained by interpolating the sounding data collected in the morning of the day of the experiment (Clements et al., 2019).

To examine effects of the horizontal mesh resolution on the fire-spread behavior, we consider four different meshes in which the vertical resolution is kept constant at $\Delta_z = 0.5$ m, and the resolution in the horizontal direction is successively refined from $\Delta = 4$ m (Case A), to $\Delta = 2$ m (Case B), to $\Delta = 1$ m (Case C), and to $\Delta = 0.5$ m (Case D). With this, our finest mesh (case D) has a total of $4.295 \times 10^9$ grid points. In addition, to examine effects of the vertical resolution, we performed two additional mesh refinement studies in which the vertical mesh resolution for Case C was refined to $\Delta_z = 0.25$ m and $\Delta_z = 0.125$ m. These two cases are denoted as Case C.1 and Case C.2, respectively. The resulting mesh configurations for all cases are summarized in Table 2.

Table 2. Mesh resolution and mesh size used for the FireFlux II simulations.

| Case | Mesh resolution $(\Delta \times \Delta \times \Delta_z)\ [m^3]$ | Mesh size $(N_x \times N_y \times N_z)$ | Total mesh size |
|------|------|------|------|
| A | $4 \times 4 \times 0.5$ | $256 \times 128 \times 2048$ | $6.711 \times 10^7$ |
| B | $2 \times 2 \times 0.5$ | $512 \times 256 \times 2048$ | $2.684 \times 10^8$ |
| C | $1 \times 1 \times 0.5$ | $1024 \times 512 \times 2048$ | $1.074 \times 10^9$ |
| C.1 | $1 \times 1 \times 0.25$ | $1024 \times 512 \times 4096$ | $2.147 \times 10^9$ |
| C.2 | $1 \times 1 \times 0.125$ | $1024 \times 512 \times 8192$ | $4.295 \times 10^9$ |
| D | $0.5 \times 0.5 \times 0.5$ | $2048 \times 1024 \times 2048$ | $4.295 \times 10^9$ |

In this study, each fire simulation was advanced over 200 s (1.7 FTT) before ignition to establish a statistically stationary flow field. To replicate the experimental ignition procedure, fire ignition was then initiated by tracing two high-temperature ignition kernels along the lateral direction at a tilting angle of $10^\circ$ (and a speed of 0.8 m/s for 110 s). The simulations ran for another 200 s to cover the full duration of the experimental observations.

# Results

In this section, we discuss the simulation results. Starting with a qualitative description of the fire behavior to generate intuition, we continue with quantitative comparisons against available measurements from the FireFlux II experiment. This is followed by statistical analysis, the examination of the dynamic behavior, and a discussion about the computational efficiency of this simulation framework.

## *Instantaneous fire-spread behavior*

Figure 1 shows instantaneous snapshots for temperature (top row) and $Q$-criterion (bottom row) of the fire spread between 120 s to 200 s after ignition. Provided by the ambient wind, a head fire dominates the fire propagation, with narrow fire flanks due to misalignment of the fire-spread rate with the ambient wind direction. In addition, the formation of a plume with counter-rotating vortex pairs is visible as indicated by the isosurface of the oxygen mass fraction of 0.1. This is a result of the large ambient wind speed, which suggests a fast dispersion of the combustion products in the downstream direction relative to the spreading of the ground fire. To illustrate the interaction between the turbulence from the atmospheric boundary layer and the fire front, we evaluate the Q-criterion. This criterion is computed as the second invariant of the velocity gradient tensor and quantifies the balance between strain and vorticity (Chong et al. 1990). The three panels in the bottom row of Fig. 1 show that large-scale vortical structures are formed at the periphery of the fire front, resulting in the formation of counter-rotating vortices that are advected into the upper atmosphere.

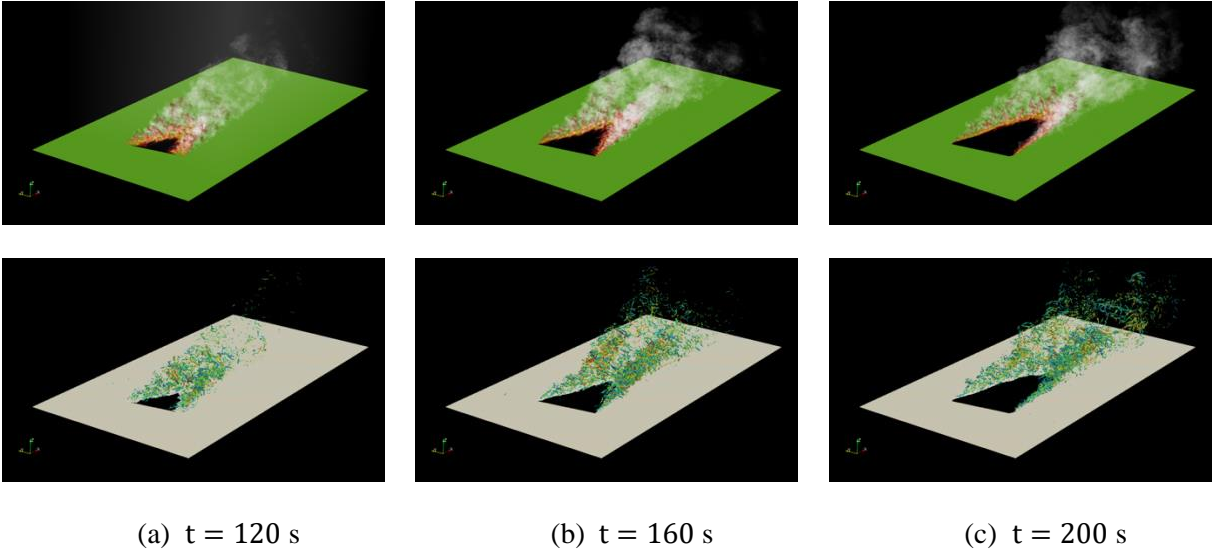

(a) t = 120 s  (b) t = 160 s  (c) t = 200 s

Figure 1. 3D visualization of the simulated fire at (a) t = 120 s , (b) t = 160 s , and (c) t = 200 s after ignition. Top row: temperature and proxy of emission plume (indicated by the isocontour of the oxygen mass fraction of 0.1); bottom row: isosurface of the $Q$-criterion (for a value of 1 s$^{-2}$) color-coded by the streamwise velocity magnitude. Results are shown for case D with 0.5 m horizontal resolution. Entire domain is shown.

The interaction between the fire and the turbulence is further illustrated in Fig. 2, showing contours of the potential temperature and vorticity in a slice perpendicular to the lateral direction. The height of the fire plume grows linearly in the upwind direction, forming an apparent angle of 50$^\circ$ with respect to the vertical direction. An average fire tilt angle of 30$^\circ$ is observed in the flame zone, and this value will be used to compute the flame length in the subsequent section. We also observe an updraft in the fire plume due to the buoyancy induced by the formation of hot products, as indicated by the velocity streamlines.

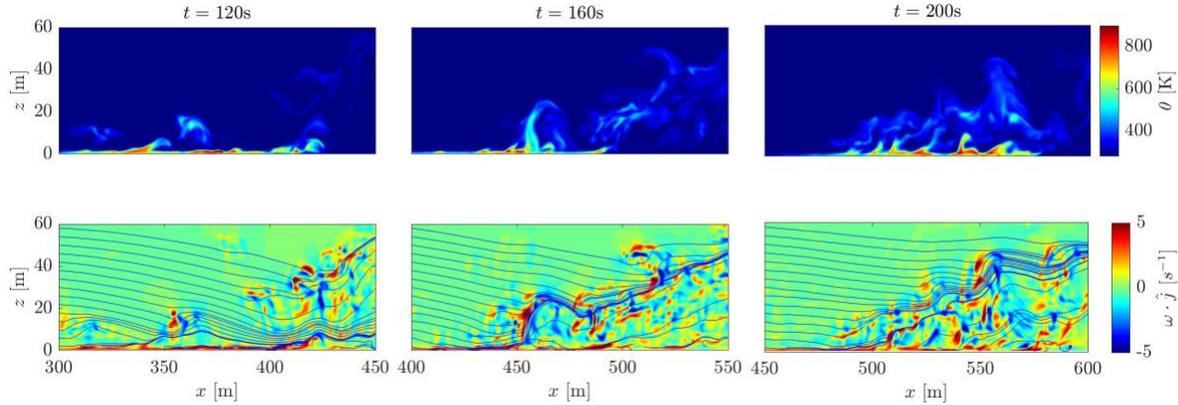

Figure 2. Potential temperature (top) and spanwise vorticity component at t = 200 s after ignition. Contours are plotted at the mid-plane normal to the lateral axis ($y$ = 250 m). Panels show a narrow region of the larger domain around the fire zone.

## Comparisons with FireFlux II experiment

Comparisons of the fire spread behavior for the six simulations of increasing spatial resolution are presented in Fig. 3, showing the gas-phase temperature at 1.5 m AGL for three different time instances, corresponding to 39 s , 44 s , and 102 s after ignition. To provide a qualitative comparison with experimental observations, we show geo-rectified IR images in the first column at the same time instances. Overall, the fire topologies predicted with different mesh resolutions show reasonable qualitative agreement with IR measurements, showing similar location and direction of the heading fire as well as the rate of spread. We note that a direct comparison with experiments is not possible due to the lack of information about the IR-image processing. Therefore in Fig. 3 we focus on a qualitative comparison with the experiment only. Results from Cases A and B show a highly diffusive flame structures with higher gas phase temperatures compared to the other cases. Between Cases C, C.1, and C.2 with a horizontal resolution of 1 m, we see similar fire structures with more corrugations in the fire zone, suggesting that the finer mesh resolution is able to better resolve the fine-scale turbulent structures and its interaction with the fire. Different vertical resolutions among these three cases do not show significant sensitivities in terms of the fire-front location. Case D has a broader flame zone than all other cases, with more fine-scale structures that are represented by regions of higher gas phase temperature inside the flame zone.

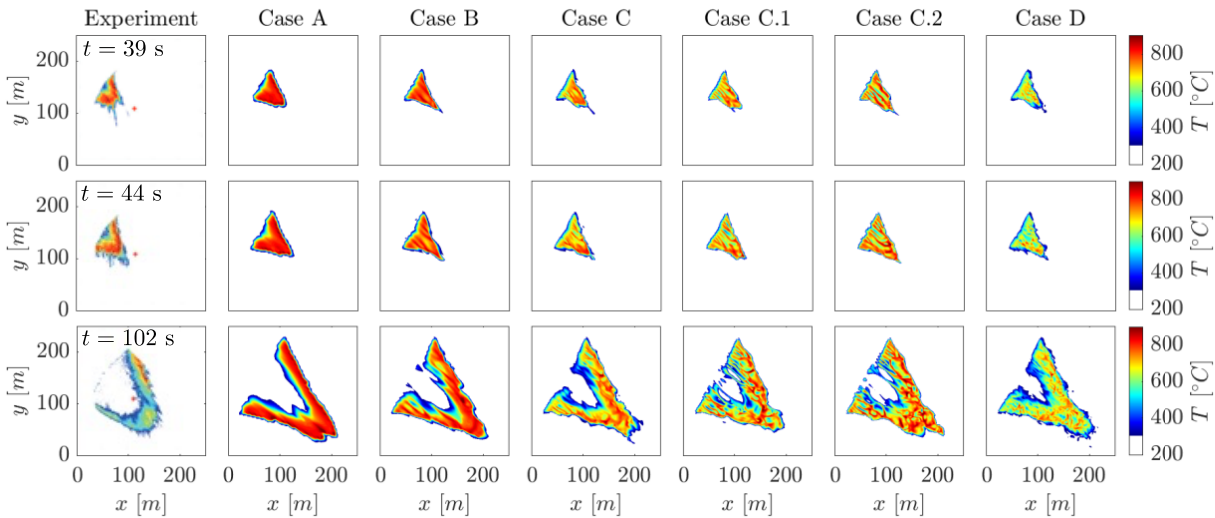

Figure 3. Qualitative comparison of gas-phase temperature field between the experiment and simulations at 1.35 m AGL. The three rows show results for 39 s (top), 44 s (middle), and 102 s (bottom) after ignition, which corresponds to 15:04:47, 15:04:52, and 15:05:50 in the experiment (Clement et al., 2019). The first column is a reference from the experiment. Columns 2 to 7 are from simulations of Cases A, B, C, C.1, C.2, and D, respectively.

The prediction of the fire-spread behavior is further examined by comparing predictions with experimentally measured isochrones for fire-front arrival. Results from this comparison are shown in Fig. 4. The black contours are obtained from the interpolated temperature field collected by the 25 HOBO sensors, and the colored contours are from the simulation with 0.5 m homogeneous spatial resolution (Case D). Overall, the simulation captures the fire-spread behavior, including the spreading rate and direction of both the fire head and flanks, reasonably well. An underprediction of the fire spreading rate is observed for the first 100 s after ignition. We attribute these differences to the lack of detailed information about the wind conditions and limitations of the fire model in representing the turbulence-combustion interaction. This is evident from the comparison of the fire spread behavior after ignition and fully developed. In the early stage, the fire is driven by the ignition and the ambient wind. As the fire becomes fully established, the fire-induced wind dominates the ambient wind, which diminishes the discrepancy between the simulation and the experiment (Clements et al., 2019).

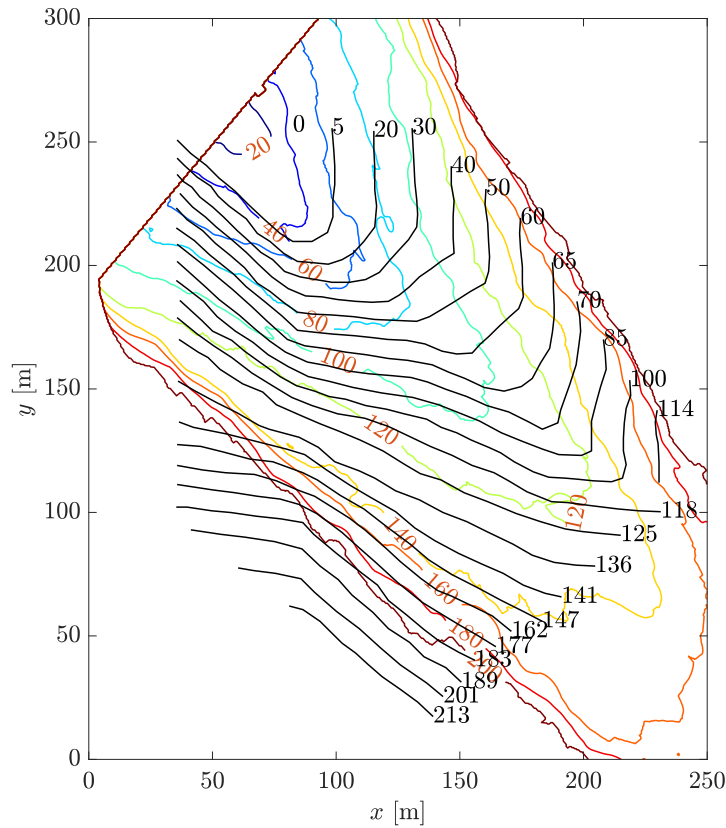

Figure 4. Comparison of isochrones for fire arrival times. Colored contours are obtained from the simulation with Case D and evaluated for the gas-phase temperature of 400℃ at 20 s intervals; black contours are from the experiment (Clements et al., 2019).

Figure 5 compares time histories of velocity and temperature at three different heights on the main tower, corresponding to 20 m, 10 m, and 5.8 m AGL. Here, only cases with different horizontal mesh resolutions (A, B, C, and D) are included for comparison. This comparison shows that the simulations capture the highly dynamic flow field and temperature intermittency, which is also observed from the experiment. A rapid increase in temperature is observed immediately after the fire front strikes the tower, which induces rapid fluctuations of velocity due to buoyancy. This behavior is captured by the simulations at all resolutions. In addition, the velocity traces show the correct responses as the fire passes by the tower. An over-prediction of the fire residence time is observed, which we attribute to uncertainties and spatial variability in the fuel load and wind conditions specified in the simulation. To improve the results and assess uncertainties to environmental variables and model sensitivities, further simulations and detailed parametric analyses are required.

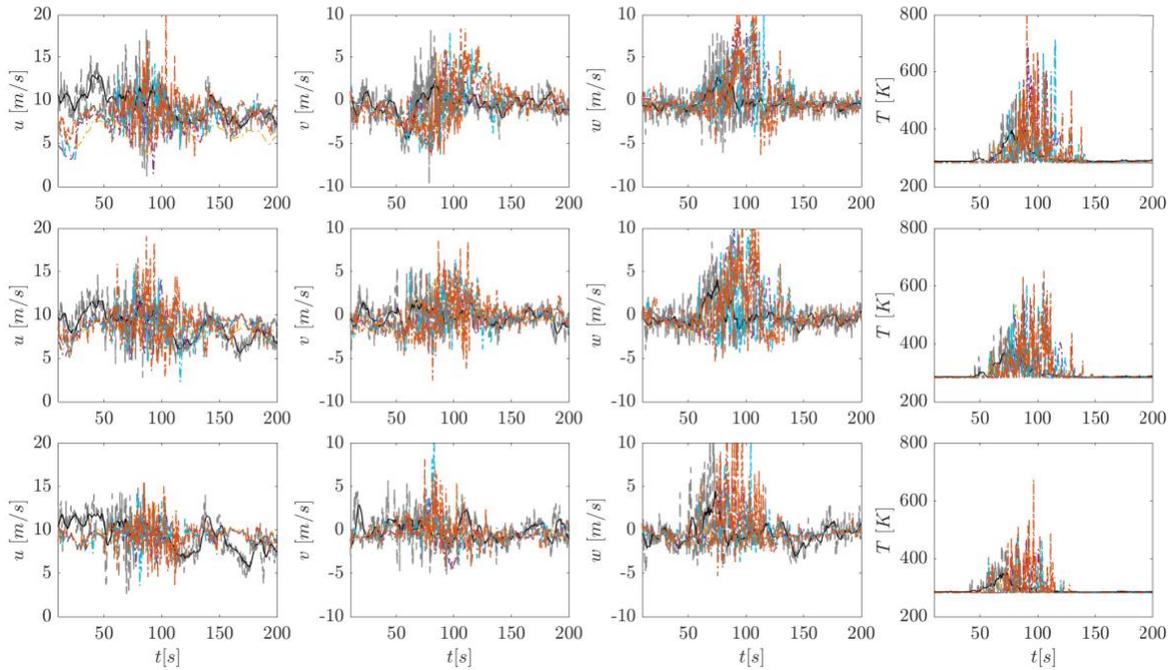

Figure 5. Time history of streamwise, lateral, and vertical velocity components and temperature at the main tower at three locations, corresponding to 20 m (top row), 10 m (middle row), and 5.8 m (bottom row) AGL. The black dashed and solid curves represent the original and filtered experimental data, respectively. Colors represent simulations with different resolutions: orange (Case A), violet (Case B), blue (Case C), red (Case D).

From the sample probe data of velocity and temperature, we compute probability density functions (PDFs) to facilitate a statistical comparison, as shown in Fig. 6. These comparisons show that the mean values and general shapes of the distributions predicted by the simulations are in reasonable agreement with experimental measurements, and the agreement improves with mesh resolutions. To quantify these results, we compared the differences of the first two moments of the streamwise velocity. These results are summarized in Table A1, showing that the largest differences in the mean-flow predictions are observed at the highest measurement location, and with largest differences observed for Case A (32.7%). The agreement improves with increasing resolution to deviations of 7.1% for Case D. In contrast, the largest differences in the velocity variance are observed for the lowest measurement locations, with deviations of 81.6% for Case A and 65% for Case D. These results suggest that while more accurate characterization of the experimentally observed wind profile can improve the mean-flow prediction, further improvements in wall models could be beneficial to improve predictions of the turbulence/fire interaction in the viscous boundary layer region.

To examine the fire intermittency, we plot the temperature PDFs in log-scale in Fig. 6. Due to the short residence time as the fire passes over the main tower, the distributions are dominated by the ambient temperature. From these comparisons, it can be seen that the simulations predict a broad temperature distribution that is skewed to lower temperatures at ambient conditions. However, it is interesting to note that the PDF for case D at 5.8 m AGL shows a bimodal distribution with a peak in temperature above 700 K, which identifies the increase of gas-phase temperature when the fire passes through the tower.

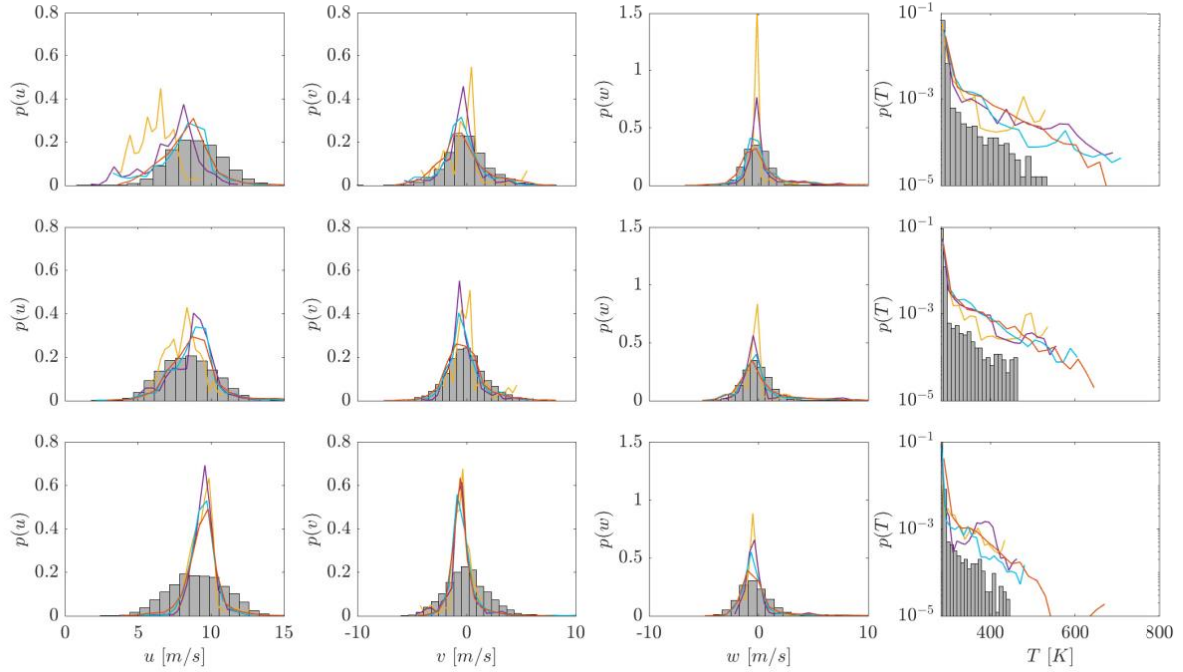

Figure 6. PDFs of streamwise, lateral, and vertical velocity components and temperature at the main tower that are 20 m (top row), 10 m (middle row), and 5.8 m (bottom row) AGL. Results shown by gray histograms are generated from the experimental data. Colors represent simulations with different resolutions: orange (Case A), violet (Case B), blue (Case C), red (Case D).

*Statistical analysis of fire behavior*

We proceed by examining global quantities. To this end, we consider the evolution of the heading fire, the fire scar, and the burning rate, as shown in Fig. 7. From results in Fig. 7(a), it can be seen that all simulations provide comparable fire-front locations that evolve with a nearly constant spreading rate of 1.6 m/s. These results are in good agreement with the experimental observations (Clement et al., 2019) as indicated by the black reference curve in Fig. 7(a). The insensitivity of these results with respect to mesh resolution suggest that the spreading rate is mainly dependent on the large-scale turbulence feedbacks between the fire and ambient wind. At the end of the simulations, the prediction in fire front deviates within 5% compared to the experiment.

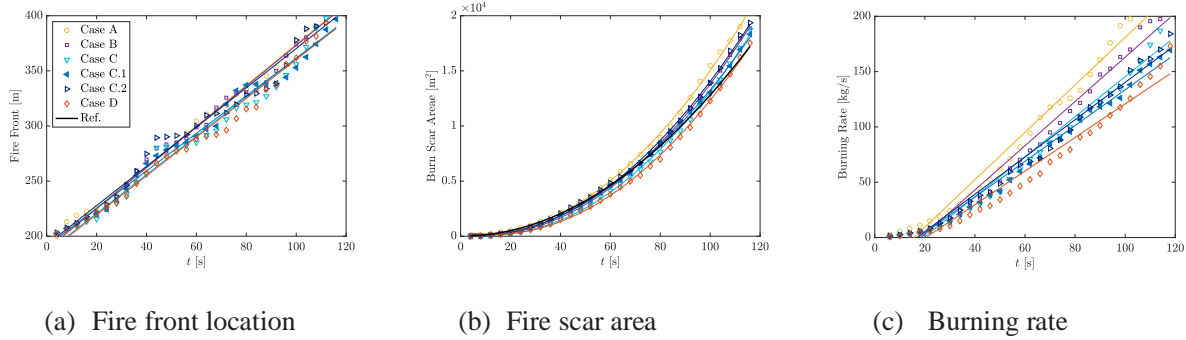

(a) Fire front location  (b) Fire scar area  (c) Burning rate

Figure 7. Fire front location, fire scar area, and burning rate for simulations with different mesh resolutions. Colors represent simulations with different resolutions: orange (Case A), violet (Case B), blue (Case C), red (Case D). Cases C, C.1, and C.2 are labeled with blue at different gray scales. The dark black curve in Fig. (a) and (b) are references computed with empirical estimation.

Figure 7(b) shows the growth of the fire scar as a function of time. All results converge to a parabolic profile over the first 120 s after ignition. Because the progressive ignition procedure with two fire torches traveling away from each other along a straight line with a speed of 1.6 m/s over the first 110 s of the fire, we can approximate the fire-front location along the baseline of ignition as $1.6t$ m. Based on the Huygens' principle of elliptic fire behavior, the location of the fire front along the wind direction can then be approximated by a triangle. The height of the triangle corresponds to the fire spread at the start of the ignition process, which can be approximated as ROS $\cdot$ t = 1.6t m. The area of the fire scar can be therefore approximated as $0.5 \cdot 1.6t \cdot 1.6t = 1.28t^2$ m$^2$, which is shown as the reference curve in Fig. 7(b). This fire behavior is well captured by all simulations, suggesting that it is largely dominated by the ignition process. In contrast, the burning rate, shown in Fig. 7(c), exhibits a more pronounced sensitivity to the horizontal mesh resolution. In particular, a reduction in the burning rate is observed with the increasing horizontal mesh resolution. With an increase of horizontal mesh resolution, small-scale turbulence structures are captured. Fire intermittency as a result of turbulence leads to a reduction in the local residence time (Viegas et al., 2021). Consequently, the solid fuel is consumed slower than it does in a fire that is subject to lower turbulence intensity. As shown in Fig. 3, at the same time after ignition, while the location of the leading edge of the fire is comparable for all simulations, the temperature fluctuations within the fire become more intense with mesh refinement. In addition, the width of the flame zone increases with increasing mesh refinement, which is consistent with longer average residence times and the presence of residual unburned fuel for longer periods of time as the fire front advances. This observation indicates that the burning rate is not fully converged in the present study, and further refinement is required.

So far, we largely examined the fire-spread behavior. To also examine the effects of the mesh resolution on the fire dynamics by buoyancy, we proceed by analyzing the ability of the mesh resolution in predicting the buoyant instabilities (Finney et al., 2015). In order to relate our analysis to experimental observations, we followed a similar procedure for evaluating the controlling parameters as done by Finney et al. (2015), describing the Strouhal number, $St = fLU^{-1}$, and Froude number, $Fr = U^2(gL)^{-1}$. Here, $f$ is the frequency, $\lambda$ is the wavelength, $L$ is the flame length, and $U$ is the ambient wind velocity magnitude. Following Finney et al. (2021), we estimate the flame length as $L = h \sec \theta$, where the flame height $h$ is determined from the simulation, and $\theta$ is the tilt angle of the fire with respect to the vertical direction. By analyzing our simulation results, the title angle was approximated as $30^\circ$. The flame height is determined by the vertical coordinates of the fire tip that is identified from the temperature contour at the pyrolysis temperature of 600 K in this study. The wavelength is measured by the average distance between the stripes in the fire front. Results are collected from 80 s to 120 s after ignition, with the flame length and wavelength obtained from cross sections aligned and normal to the mean wind direction, respectively, and are shown in Fig. 8(a). A wide range of distribution of flame lengths is observed for all simulations, which is due to the unsteadiness and variability of the flame height in response to turbulence. Figure 8(a) shows that the buoyancy wavelength remains approximately consistent for horizontal resolutions better than 1 m, showing little sensitivity to vertical resolution (as illustrated by Cases C, C1. and C2 having vertical resolution of 0.5, 0.25, and 0.125 m, respectively).

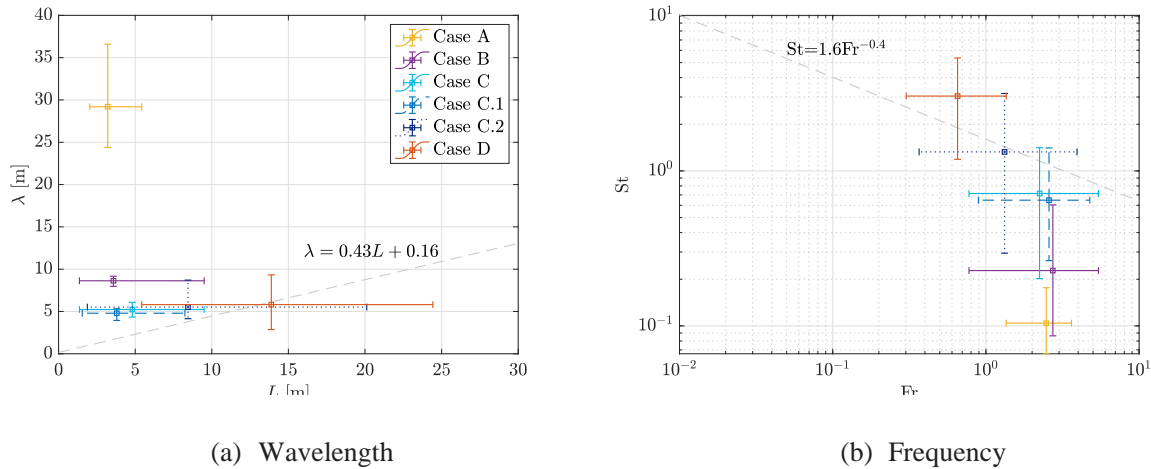

(a) Wavelength

(b) Frequency

Figure 8. Characteristic length and time scales of the buoyant fire dynamics between 80 s and 120 s after ignition.

Figure 8(b) shows the fire's temporal response at the main tower 20 m AGL. The frequency $f$ is computed as the inverse of the average period of temperature fluctuation at a specific location. These results show that the frequency and flame height (expressed in terms of flame length in our analysis) exhibit more pronounced sensitivities to the mesh resolution. As a reference, we also include experimentally reported empirical correlations by the black dashed line to guide to reader. We believe that the results presented in this analysis are useful to assess physical models in accurately predicting the buoyant flame dynamics and plume physics, as well as examining sensitivities to capturing these instabilities.

*Computational performance*

To evaluate the performance of the solver and assess the scalability towards enabling large-scale wildland fire simulations, we performed benchmark simulations using 128 TPU cores for the cases A, B, C, and D, and an additional case that has a horizontal mesh resolution of 8 m with a per-core mesh size of $128 \times 64 \times 16$. The results are summarized in Fig. 9. For each level of mesh refinement, the number of grid points per core is quadrupled, which provides a maximum number of grid points per core close to 40 million with a memory utilization of 12 GiB/core. From Fig. 9, for a fixed computational resource, we see a nearly linear speedup with mesh size except for the case with the smallest number of grid points per processor. We attribute this degradation in performance to the underutilization of the MXU compute kernel, which is optimized for performing matrix multiplications of size $128 \times 128$ per instruction. In addition, in our previous study (Wang et al., 2022a), we have demonstrated that this simulation framework provides linear weak and strong scalability to the full TPU pod.

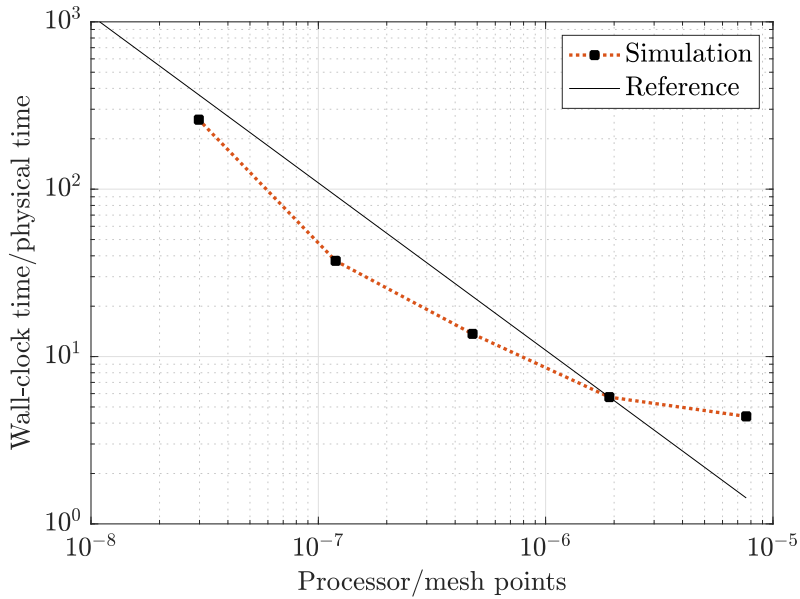

Figure 9. Simulation wall-clock time normalized by physical time as a function of the number of processors allocated per mesh point.

With relevance to performing large-scale wildland fire simulations, we estimate the largest problem size that we are able to simulate using the currently available TPU architectures with 8192 cores (TPUv4 pod). Considering a vertical mesh resolution of 1 m in case B and scaling it to a full TPU pod, we estimate being able to simulate domains of 100 km$^2$ in size. This is comparable to the burned area of the Northern California Tubbs fire (Martinez et al., 2017), making it feasible to examine the first three hours during the extreme fire development within 1.3 days of run time (Wang et al., 2022b).

## Conclusions

In this work, we developed a simulation framework utilizing TensorFlow programming paradigm and TPU hardware, to enable large-scale high-resolution simulations of wildland fires. This simulation framework adopts a quasi-physical model for the representation of the combustion, which is coupled to a large-eddy simulation for the representation of the atmospheric flow field. We performed simulations of the FireFlux II configuration with increasing levels of mesh resolutions, ranging from 4 m to 0.5 m horizontally and 0.5 m to 0.125 m vertically, and evaluated results against the experiment. The predictions for velocity and temperature at the main tower are compared with available experimental data. Both the temperature projection AGL and the isochrone contours are in reasonable agreement with the experimental observations after the fire is fully established, which demonstrates that the simulation predicts the fire spread in a reasonable manner. Discrepancies at the early stage of the fire after ignition can be attributed to uncertainties in the wind and fuel conditions as well as limitations of the combustion model in representing the combustion processes and turbulence interaction. Results from the mesh-resolution analysis indicate that predictions of intermittent fire behavior, buoyancy-driven dynamics, and processes that are affected by small-scale turbulent motion benefit from improved mesh refinement.

Results from the scaling analysis show a close to linear scalability of the simulation framework, and conservative estimates indicate that it becomes feasible to perform large-scale simulations of wildland fire scenarios comparable to the Tubbs fire at a spatial resolution of 1 m at acceptable computation time. As such, the capability of generating such high-resolution simulation results for meso-scale wildland fires makes the simulations useful for scientific discovery, forensic analysis, and fire management under realistic conditions and spatio-temporal resolutions.

## References

M.Abadi, A. Agarwal, P. Barham, E. Brevdo, Z. Chen, C. Citro, G. S. Corrado, A. Davis, J. Dean, M. Devin, S. Ghemawat, I. Goodfellow, A. Harp, G. Irving, M. Isard, Y. Jia, R. Jozefowicz, L. Kaiser, M. Kudlur, J. Levenberg, D. Man´e, R. Monga, S. Moore, D. Murray, C. Olah, M. Schuster, J. Shlens, B. Steiner, I. Sutskever, K. Talwar, P. Tucker, V. Vanhoucke, V. Vasudevan, F. Vi´egas, O. Vinyals, P. Warden, M. Wattenberg, M. Wicke, Y. Yu, X. Zheng, TensorFlow: Large-scale machine learning on heterogeneous systems, 2015.

Abatzoglou, J. T., & Williams, A. P. (2016). Impact of anthropogenic climate change on wildfire across western US forests. *Proceedings of the National Academy of Sciences* , *113*(42), 11770–11775.

Andrews, P. L. (2009). *BehavePlus fire modeling system, version 5.0: Variables*. General Technical Report RMRS-GTR-213WWW Revised, Fort Collins, CO, United States Department of Agriculture, Forest Service, Rocky Mountain Research Station.

Balbi, J. H., Morandini, F., Silvani, X., Filippi, J. B., & Rinieri, F. (2009). A physical model for wildland fires. *Combustion and Flame*, *156*(12), 2217–2230.

Bakhshaii, A., & Johnson, E. A. (2019). A review of a new generation of wildfire–atmosphere modeling. In *Canadian Journal of Forest Research,* 49(6), 565–574.

Burke, M., Driscoll, A., Heft-Neal, S., Xue, J., Burney, J., & Wara, M. (2021). The changing risk and burden of wildfire in the United States. *Proceedings of the National Academy of Sciences,* 118(2)*,* e2011048118.

Chong, M. S., Perry, A. E. & Cantwell, B. J. (1990). A general classification of three-dimensional flow fields. *Physics of Fluids*, *2*, 765–777.

Clements, C. B., Kochanski, A. K., Seto, D., Davis, B., Camacho, C., Lareau, N. P., Contezac, J., Restaino, J., Heilman, W. E., Krueger, S. K., Butler, B., Ottmar, R. D., Vihnanek, R., Flynn, J., Filippi, J.-B., Barboni, T., Hall, D. E., Mandel, J., Jenkins, M. A., O'Brien, O., Hornsby, B., & Teske, C. (2019). The FireFlux II experiment: A model-guided field experiment to improve understanding of fire–atmosphere interactions and fire spread. *International Journal of Wildland Fire*, *28*(4), 308–326.

Coen, J. (2013). Modeling wildland fires: A description of the Coupled Atmosphere-Wildland Fire Environment model (CAWFE) (No. NCAR/TN-500+STR). NCAR Earth System Laboratory.

Coen, J. L., Cameron, M., Michalakes, J., Patton, E. G., Riggan, P. J., & Yedinak, K. M. (2013). WRF-Fire: Coupled Weather–Wildland Fire Modeling with the Weather Research and Forecasting Model. *Journal of Applied Meteorology and Climatology*, *52*(1), 16–38.

Coen, J. L., Schroeder, W., Conway, S., & Tarnay, L. (2020). Computational modeling of extreme

wildland fire events: A synthesis of scientific understanding with applications to forecasting, land management, and firefighter safety. *Journal of Computational Science*, *45*, 101152.

de Gennaro, M., Billaud, Y., Pizzo, Y., Garivait, S., Loraud, J.-C., El Hajj, M., & Porterie, B. (2017). Real-time wildland fire spread modeling using tabulated flame properties. *Fire Safety Journal*, *91*, 872–881.

Filippi, J.-B., Bosseur, F., Pialat, X., Santoni, P.-A., Strada, S., & Mari, C. (2011). Simulation of coupled fire/atmosphere interaction with the MesoNH-ForeFire models. *Journal of Combustion*, 2011, 540390.

Finney, M. A. (2004). FARSITE: fire area simulator-model development and evaluation. Research Paper RMRS-RP-4. *US Department of Agriculture, Forest Service, Rocky Mountain Research Station, Ogden, Utah, USA*.

Finney, M. A., Cohen, J. D., Forthofer, J. M., McAllister, S. S., Gollner, M. J., Gorham, D. J., Saito, K., Akafuah, N. K., Adam, B. A., & English, J. D. (2015). Role of buoyant flame dynamics in wildfire spread. *Proceedings of the National Academy of Sciences of the United States of America*, *112*(32), 9833–9838.

Hessburg, P. F., Prichard, S. J., Hagmann, R. K., Povak, N. A., & Lake, F. K. (2021). Wildfire and climate change adaptation of western North American forests: A case for intentional management. *Ecological Applications: A Publication of the Ecological Society of America*, *31*(8), e02432.

Jolly, W. M., Cochrane, M. A., Freeborn, P. H., Holden, Z. A., Brown, T. J., Williamson, G. J., & Bowman, D. M. J. S. (2015). Climate-induced variations in global wildfire danger from 1979 to 2013. *Nature Communications*, *6*, 7537.

Jouppi, N. P., Hyun Yoon, D., Ashcraft, M., Gottscho, M., Jablin, T. B., Kurian, G., Laudon, J., Li, S., Ma, P., Ma, X., Norrie, T., Patil, N., Prasad, S., Young, C., Zhou, Z., & Patterson, D. (2021). ten lessons from three generations shaped Google's TPUv4: Industrial product. *2021 ACM/IEEE 48th Annual International Symposium on Computer Architecture (ISCA)*, 1–14.

Klemp, J. B., & Lilly, D. K. (1978). Numerical simulation of hydrostatic mountain waves. *Journal of the*

*Atmospheric Sciences*, 35(1), 78–107.

Leutbecher, M., & Palmer, T. N. (2008). Ensemble forecasting. *Journal of Computational Physics*, *227*(7), 3515–3539.

Linn, R. R. (1997). *A Transport Model for Prediction of Wildfire Behavior (No. LA-13334-T)*, Ph.D. thesis. Los Alamos National Lab., NM.

Linn, R., Reisner, J., Colman, J. J., & Winterkamp, J. (2002). Studying wildfire behavior using FIRETEC. *International Journal of Wildland Fire,* 11(4), 233-246.

Linn, R. R., Sieg, C. H., Hoffman, C. M., Winterkamp, J. L., & McMillin, J. D. (2013). Modeling wind fields and fire propagation following bark beetle outbreaks in spatially-heterogeneous pinyon-juniper woodland fuel complexes. *Agricultural and Forest Meteorology*, 173, 139–153.

Linn, R. R., & Cunningham, P. (2005). Numerical simulations of grass fires using a coupled atmosphere–fire model: Basic fire behavior and dependence on wind speed. *Journal of Geophysical Research*, *110*(D13), 287.

Liu, Y., Kochanski, A. Baker, K. R., Mell, W., Linn, R., Paugam, R., Mandel, J., Fournier, A., Jenkins, M. A., Goodrick, S., Achtemeier, G., Zhao, F., Ottmar, R., French, N. H. F., Larkin, N., Brown, T., Hudak, A., Dickinson, M., Potter, B., Clements, C., Urbanski, S., Prichard, S., Watts, A., McNamara, D. (2019). Fire behaviour and smoke modelling: model improvement and measurement needs for next-generation smoke research and forecasting systems. *International Journal of Wildland Fire.* 28, 570-588.

Lilly, D. K. (1962). On the numerical simulation of buoyant convection. *Tellus*, *14*(2), 148–172.

Mandel, J., Beezley, J. D., & Kochanski, A. K. (2011). Coupled atmosphere-wildland fire modeling with WRF 3.3 and SFIRE 2011. *Geoscientific Model Development*, *4*(3), 591–610.

Martinez, J., Bergland, V., Franklin, M., Frits, M., Lohse, S., Roath, G. and Thompson, M., Investigation Report, 17CALNU010045, California Department of Forestry and Fire Protection, Sonoma-Lake Napa Unit, October 8, 2017.

Moeng, C.-H. (1984). A large-eddy-simulation model for the study of planetary boundary-layer turbulence. *Journal of the Atmospheric Sciences*, 41(13), 2052–2062.

Moinuddin, K. A. M., Sutherland, D., and Mell, W. (2018). Simulation study of grass fire using a physics-based model: striving towards numerical rigour and the effect of grass height on the rate of spread, *International Journal of Wildland Fire*, 27(12), 800-814.

Morvan, D., Accary, G., Meradji, S., Frangieh, N., Bessonov, O. (2018). A 3D physical model to study the behavior of vegetation fires at laboratory scale, *Fire Safety Journal*, 101, 39–52.

McGrattan, K. B., McDermott, R., Vanella, M., Hostikka, S., & Floyd, J. (2006). *Fire Dynamics Simulator Technical Reference Guide* (No. 1018-2; 6th ed.). National Institute of Standards and Technology.

Ntaimo, L., Hu, X., & Sun, Y. (2008). DEVS-FIRE: Towards an integrated simulation environment for surface wildfire spread and containment. *Simulation*, *84*(4), 137–155.

Parks, S. A., & Abatzoglou, J. T. (2020). Warmer and drier fire seasons contribute to increases in area burned at high severity in western us forests from 1985 to 2017. *Geophysical Research Letters*, 47(22), e2020GL089858.

Porterie, B., Morvan, D., Loraud, J. C., & Larini, M. (2000). Firespread through fuel beds: Modeling of wind-aided fires and induced hydrodynamics. *Physics of Fluids*, *12*(7), 1762–1782.

Rothermel, R. C. (1972). A mathematical model for predicting fire spread in wildland fuels. *Research Paper INT-115*, Intermountain Forest & Range Experiment Station, Forest Service, US Department of Agriculture.

Siebesma, A. P., Pier Siebesma, A., Bretherton, C. S., Brown, A., Chlond, A., Cuxart, J., Duynkerke, P. G., Jiang, H., Khairoutdinov, M., Lewellen, D., Moeng, C.-H., Sanchez, E., Stevens, B., & Stevens, D. E. (2003). A large eddy simulation intercomparison study of shallow cumulus convection. *Journal of the Atmospheric Science*, *60*(10), 1201-1219.

Stoll, R., & Porté-Agel, F. (2006). Dynamic subgrid-scale models for momentum and scalar fluxes in large-eddy simulations of neutrally stratified atmospheric boundary layers over heterogeneous terrain. *Water Resources Research*, *42*(1), W01409.

Sullivan, A. L. (2009a). Wildland surface fire spread modelling, 1990–2007. 1: Physical and quasi-physical models. *International Journal of Wildland Fire,* 18, 349–368.

Sullivan, A. L. (2009b). Wildland surface fire spread modelling, 1990–2007. 2: Empirical and quasi-empirical models. *International Journal of Wildland Fire,* 18, 369–386.

Sullivan, A. L. (2009c). Wildland surface fire spread modelling, 1990–2007. 3: Simulation and mathematical analogue models. *International Journal of Wildland Fire,* 18, 387–403.

Thomas, D., Butry, D., Gilbert, S., Webb, D., & Fung, J. (2017). *The costs and losses of wildfires: a literature survey*. NIST Special Publication 1215, National Institute of Standards and Technology

Viegas, D. X. F., Raposo, J. R. N., Ribeiro, C. F. M., Reis, L., Abouali, A., Ribeiro, L. M., & Viegas, C. X. P. (2022). On the intermittent nature of forest fire spread – Part 2. *International Journal of Wildland Fire*, 31(10), 967–981.

Wang, Q., Ihme, M., Chen, Y.-F., & Anderson, J. (2022a). A TensorFlow simulation framework for scientific computing of fluid flows on tensor processing units. *Computer Physics Communications,* 274, 108292.

Wang, Q., Ihme, M., Chen, Y.-F., Yang, V., Sha, F., & Anderson, J. (2022b). Towards real-time predictions of large-scale wildfire scenarios using a fully coupled atmosphere-fire physical modelling framework. In V. D. X. R. Mario (Ed.), *Advances in Forest Fire Research* (pp. 415–421).

Wang, Y., Chatterjee, P., & de Ris, J. L. (2011). Large eddy simulation of fire plumes. *Proceedings of the Combustion Institute*, 33(2), 2473–2480.

Westerling, A. L. (2016). Increasing western US forest wildfire activity: Sensitivity to changes in the timing of spring. *Philosophical Transactions of the Royal Society of London. Series B, Biological Sciences*, 371, 20150178.

Westerling, A. L., Hidalgo, H. G., Cayan, D. R., & Swetnam, T. W. (2006). Warming and earlier spring increase western U.S. Forest wildfire activity. *Science*, 313(5789), 940–943.

Wilson, R. (1980). Reformulation of forest fire spread equations in SI units (Vol. 292). Department of Agriculture, Forest Service, Intermountain Forest and Range Experiment Station.

Yang, X. I. A., Chen, P. E. S., Hu, R., & Abkar, M. (2022). Logarithmic-linear law of the streamwise velocity variance in stably stratified boundary layers. *Boundary-Layer Meteorology*, 183(2), 199–213.

## Appendix A: Governing equations

The spatial and temporal evolution of the wildland fire through the combustion of solid fuel and the coupling to the atmospheric flow is described by a two-phase model (Linn, 1997). In this model, the gas-phase is described by the solution of the Favre-filtered conservation equations for mass, momentum, oxygen-mass fraction, and potential temperature as:

$$\partial_t \bar{\rho} + \nabla \cdot (\bar{\rho} \, \widetilde{\boldsymbol{u}}) = S_\rho, \tag{1}$$

$$\partial_t (\bar{\rho} \widetilde{\boldsymbol{u}}) + \nabla \cdot (\bar{\rho} \widetilde{\boldsymbol{u}} \otimes \widetilde{\boldsymbol{u}}) = -\nabla \overline{p_d} + \nabla \cdot \bar{\tau} + [\bar{\rho} - \rho(z)] g \widehat{\boldsymbol{k}}_z + \boldsymbol{f}_D + \boldsymbol{f}_C, \tag{2}$$

$$\partial_t (\bar{\rho} \widetilde{Y_O}) + \nabla \cdot (\bar{\rho} \widetilde{\boldsymbol{u}} \widetilde{Y_O}) = \nabla \cdot \overline{J_O} + \bar{\rho} \widetilde{\dot{\omega}_O}, \tag{3}$$

$$\partial_t (\bar{\rho} \bar{\theta}) + \nabla \cdot (\bar{\rho} \widetilde{\boldsymbol{u}} \bar{\theta}) = \nabla \cdot \overline{\boldsymbol{q}} + \frac{\bar{\rho} \bar{\theta}}{c_p \bar{T}} \left[ h a_v (T_s - \bar{T}) + \dot{q}_r + (1 - \Theta) H_f \widetilde{\dot{\omega}} \right], \tag{4}$$

where $\rho$ is the density, $\boldsymbol{u}$ is the velocity vector, $p_d$ is the hydrodynamic pressure, $\tau$ is the shear stress tensor, $g$ is the gravitational acceleration constant, $\widehat{\boldsymbol{k}}_z$ is the unit vector along the gravitational direction, $f_D = -\bar{\rho} c_d a_v |\widetilde{\boldsymbol{u}}| \widetilde{\boldsymbol{u}}$ is the drag force due to surface vegetation and the drag coefficient is set to $c_d = 0.01$ (Linn & Cunningham, 2005), $\boldsymbol{f}_C = f \widehat{\boldsymbol{k}}_z \times \bar{\rho} (\widetilde{\boldsymbol{u}} - \boldsymbol{U}_\infty)$ is the Coriolis force (Siebesma et al., 2003) with $f = -2\Omega \sin \psi$ being the Coriolis coefficient, $\Omega$ being the rotation rate of the earth and $\psi$ being the latitude, $Y_O$, $\boldsymbol{j}_O$, and $\dot{\omega}_O$ are the mass fraction, species diffusion, and source term of the oxidizer $O$, $\theta$ is the potential temperature, $\boldsymbol{q}$ is the heat flux vector, $T$ is the gas-phase temperature, and $H_f$ is the heat of combustion. The Favre-filtered value for a generic variable $\phi$ is defined with the tilde notation as $\tilde{\phi} = \overline{\rho \phi} / \bar{\rho}$, where the overbar denotes the Reynolds filtering. The heat exchange with the solid fuel is accounted for by the convective heat transfer, with $h$ being the convective heat transfer coefficient, and $a_v$ being the bulk fuel area-to-volume ratio. $\Theta = 1 - \rho_f / \rho_{f,0}$ is the fraction of the heat release that contributes to the increase of the solid phase temperature (Linn, 1997). The shear stress tensor, combining molecular and turbulent transport, is computed as $\bar{\tau} = 2(\mu + \mu_t) \tilde{S}$, with $\tilde{S} = [\nabla \widetilde{\boldsymbol{u}} + (\nabla \widetilde{\boldsymbol{u}})^T]/2 + (\nabla \cdot \widetilde{\boldsymbol{u}} I)/3$ being the strain rate tensor and $\mu$ the dynamic viscosity, which is related to the kinematic viscosity with $\nu = \mu / \rho = 10^{-5}$ m²/s. In this study, the Smagorinsky model is used to compute the eddy viscosity, which is $\mu_t = \bar{\rho} \nu_t$ where $\nu_t =$ $(C_s\Delta)^2|\tilde{S}|$, with a constant Smagorinsky coefficient $C_s = 0.18$. The diffusive flux for oxygen is $\overline{J_O} = \bar{\rho}(\alpha + \alpha_t)\nabla\widetilde{Y_O}$, and the heat flux is $\overline{q} = \bar{\rho}(\lambda + \lambda_t)\nabla\tilde{\theta}$, where turbulent diffusivity and conductivity are computed as $\alpha_t = \nu_t/\text{Sc}_t$ and $\lambda_t = \nu_t/\text{Pr}_t$, with $\text{Sc}_t$ and $\text{Pr}_t$ being the turbulent Schmidt number for oxygen and the turbulent Prandtl number, respectively.

Using a low-Mach number approximation, we decompose the pressure into hydrostatic and hydrodynamic components, $p = p(z) + p_d$, with the hydrostatic pressure computed as $p(z) = p_0\left[1 - gz/(c_p\theta_\infty)\right]^{1/\kappa}$, where $\theta_\infty$ is the potential temperature of the ambient air that is assumed to be a constant, $p_0$ is the atmospheric pressure on the ground level, and $\kappa = R/c_p$ is the ratio between the gas constant and specific heat. The temperature is computed as $\bar{T} = \tilde{\theta}[p(z)/p_0]^\kappa$.

The radiation source term is modeled by a gray-gas model as $\dot{q}_r = -(\sigma k/\zeta)(T_\infty^4 - \bar{T}^4)$, with $\sigma$ being the Boltzman coefficient, $k$ being a coefficient that models the turbulence-radiation interaction and is set to 1 for a balanced interaction (Linn, 1997), $T_\infty$ being the ambient temperature, and $\zeta$ being a characteristic length scale of the fuel elements that is set to 0.5 m for tall grass.

The gas-phase dynamics is coupled to the solid-fuel pyrolysis and the gas-phase reaction. In the present work, we describe the combustion by a one-step global reaction that represents both the pyrolysis of solid fuel and the reaction in the gas phase (Linn, 1997):

$$\nu_F F + \nu_O O \rightarrow \nu_P P, \tag{5}$$

where $\nu_F$, $\nu_O$, and $\nu_P$ are the stoichiometric coefficients of the fuel $F$ in the solid phase, the oxygen $O$ and the combustion products $P$ in the gas phase, respectively. Denoting $N$ as the nitrogen that is treated as an inert species in the current formulation, the Favre filtered species mass fractions $\widetilde{Y_\alpha}$ for species $\alpha \in \{F, O, N, P\}$ satisfy

$$\sum_{\alpha\in\{F,O,N,P\}}\widetilde{Y_\alpha} = 1. \tag{6}$$

In the present model, pyrolysis and gas-phase combustion are combined, which is represented through the following reaction rate:

$$\widetilde{\omega} = \frac{c_F \rho_F \bar{\rho} \widetilde{Y_O} \nu_t \Psi_S \lambda_{OF}}{\rho_\infty s_x^2},$$ (7)

where $c_F = 0.5$ is an empirical scaling coefficient (Linn, 1997), $\rho_F$ is the bulk density of the fuel that is defined as the ratio between the fuel load and the height of the fuel, $\bar{\rho}$ is the filtered gas phase density, $\rho_\infty = 1$ kg/m$^3$ is the reference density, $\nu_t$ is the turbulent diffusivity, and $s_x = 0.05$ m is an empirical coefficient to parameterize the characteristic turbulence scale. In this model, $\lambda_{OF}$ is introduced as a coefficient that maximizes the reaction rate, which is formulated as:

$$\lambda_{OF} = \frac{\rho_F \bar{\rho} \widetilde{Y_O}}{(\rho_F/\nu_F + \bar{\rho} \widetilde{Y_O}/\nu_O)^2}.$$ (8)

The linear temperature function $\Psi_S$ in Eq. (7) represents the ignited volume fraction, which is modeled as a function of the gas phase temperature $\widetilde{T}$:

$$\Psi_S = \max\left(\min\left(\frac{\widetilde{T} - T_{min}}{T_{max}}, 1\right), 0\right),$$ (9)

where $T_{min} = 300$ K and $T_{max} = 400$ K. The fuel moisture is modeled as the bulk density of liquid water, $\rho_W$. The rate of evaporation is modeled as a function of the gas phase temperature, which is similar to the reaction source term and takes the form:

$$\dot{s}_W = -\frac{\rho_{W,0}}{\Delta t} \max(\Psi_W - \max(\Psi_{W,old}), 0),$$ (10)

where $\rho_{W,0}$ is the initial bulk density of moisture, $\Psi_W = \max\left(\min\left(\frac{\widetilde{T} - T_W}{\Delta T}, 1\right), 0\right)$ is the amount of water to be evaporated with $T_w = 310$ K and $\Delta T = 126$ K, and $\max(\Psi_{W,\text{old}})$ is the maximum amount of water that has been evaporated.

The solid states, including the fuel load, the moisture content, and the temperature of the fuel, are modeled with the following ordinary differential equations (Linn & Cunningham, 2005):

$$d_t \rho_F = -\nu_F \widetilde{\omega},$$ (11)

$$d_t \rho_F = \dot{s}_W,$$ (12)

$$\left(c_{p,F} \rho_F + c_{p,W} \rho_W\right) d_t T_s = \dot{q}_{r,s} + h a_\nu (\bar{T} - T_s) + (H_W + c_{p,W} T_{\text{vap}}) \dot{s}_W + (\Theta H_f - c_{p,F} T_{\text{pyr}} \nu_F) \widetilde{\omega},$$ (13)

where $c_{p,F}$ and $c_{p,W}$ are the specific heat of the fuel and liquid water, respectively, $H_W$ is the heat of vaporization, $T_{\text{vap}}$ is the temperature of vaporization, and $T_{\text{pyr}}$ is the temperature at which the solid fuel begins to pyrolyze. Note that the source terms due to combustion and evaporation are incorporated in the gas phase equations through a Lie splitting scheme (Trotter, 1959), where the hydrodynamics are integrated following the source terms.

As a result of the combustion and the associated phase exchange, an additional source term is required in Eq. (1) for mass conservation, which is evaluated as $S_\rho = v_F \widetilde{\dot{\omega}} - \dot{s}_W$. The gas is assumed to be thermodynamically perfect, hence the equation of states is modeled by the ideal gas law:

$$p(z) = \bar{\rho} \tilde{R} \tilde{T},$$

where $\tilde{R}$ is the gas constant, which is a function of the filtered species mass fractions $\widetilde{Y_\alpha}$.

## Appendix B: Turbulent boundary layer structure

The inflow boundary condition is created from an independent simulation of a neutrally stratified boundary layer. The same computational domain as the fire simulation, which is $1000 \times 500 \times 1200 \text{ m}^3$, is used in this boundary layer simulation, with the inflow-outflow boundary conditions along the streamwise direction being replaced by a periodic boundary condition. To drive this flow, we applied a Coriolis force at a latitude of $31.5°$N that corresponds to the location of Texas where the experiment was conducted, which provides a Coriolis coefficient of $f = 7.6 \times 10^{-5} \text{ s}^{-1}$ in Eq. (2). A surface roughness of $z_0 = 0.15$ m is used to model the surface shear stress with the Monin-Obukhov similarity theory. The inflow profiles are collected after 100 flow-through times when the turbulence is fully established and the boundary layer statistics are converged. Based on the result and parameters we specified, we compute the wall shear stress as $\tau_{i3} = [\kappa(z)|\tilde{\mathbf{u}}(z)|/\ln(z/z_0)]^2 (\tilde{u}_i(z)/|\tilde{\mathbf{u}}(z)|)$, where $\kappa = 0.4$ is the von Karman constant, and $z = 5$ m is the height of the viscous sublayer. The friction velocity in this flow is found to be $u^* = (\tau_{13}^2 + \tau_{23}^2)^{1/4} = 0.36$ m/s (Moeng, 1984). Given that the boundary layer height in this study is $\delta = l_z = 1200$ m and the kinematic viscosity is $\nu = 10^{-5} \text{ m}^2/\text{s}$, the Reynolds number is determined as $\text{Re}_\tau = u_\tau \delta / \nu = 4.32 \times 10^7$ (Yang et al., 2022).

The boundary layer structure is presented in Fig. A1, showing (left) the instantaneous axial velocity field, evaluated at a horizontal location as a function of $z$ and $t$, and (right) the mean velocity profile in inner-scale variables. The viscous sublayer and the logarithmic region are captured below 200 m, providing the representation of the hydrodynamics in regions where the fire-atmosphere interaction is most intense.

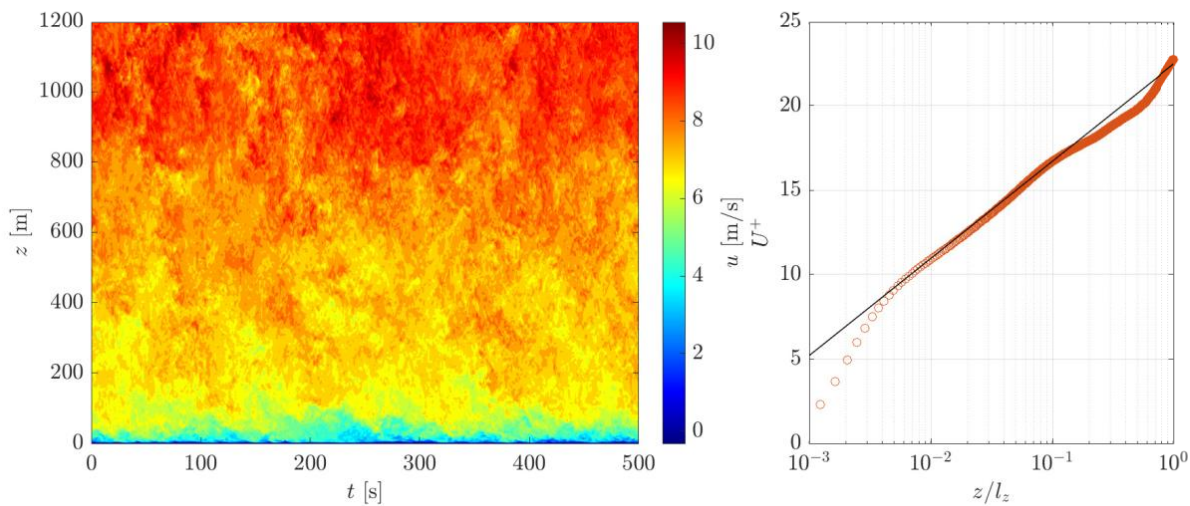

Figure A1. The inflow velocity collected from an independent simulation in the same computational domain without fire. The contour shows the time history of the instantaneous streamwise velocity in the middle of the inflow plane. The mean velocity profile shown on the right is compared with the log law of the wall. The friction velocity is $u_\tau = 0.36$ m/s, and the corresponding Reynolds number is $\text{Re}_\tau = u_\tau l_z / \nu = 4.32 \times 10^7$.

Table A1. The mean and variance of the streamwise velocity (with deviations to experiments shown in brackets) at the main tower, computed from the PDFs, shown in Fig. 6.

| Height [m] | Experiment | | Case A | | Case B | | Case C | | Case D | |
|---|---|---|---|---|---|---|---|---|---|---|
| | Mean | Var. | Mean | Var. | Mean | Var. | Mean | Var. | Mean | Var. |
| 20 | 9.0 | 3.154 | 6.04 (32.7%) | 1.42 (55.0%) | 7.30 (18.7%) | 3.21 (1.9%) | 8.00 (10.9%) | 3.14 (0.4%) | 8.34 (7.1%) | 3.12 (1.2%) |
| 10 | 8.42 | 3.255 | 8.0 (5.2%) | 1.29 (60.5%) | 8.70 (3.2%) | 1.98 (39.1%) | 8.69 (3.2%) | 2.37 (27.2%) | 8.8 (4.6%) | 3.22 (1.2%) |
| 5.8 | 9.07 | 3.642 | 9.38 (3.4%) | 0.67 (81.6%) | 9.43 (3.9%) | 0.59 (85.2%) | 9.42 (3.8%) | 0.87 (76.1%) | 9.4 (3.7%) | 1.28 (65.0%) |